\documentclass[lettersize,journal]{IEEEtran}
\usepackage{amsmath,amsfonts}
\usepackage{algorithmic}
\usepackage{algorithm}
\usepackage{array}
\usepackage[caption=false,font=footnotesize,labelfont=rm,textfont=rm]{subfig}
\usepackage{textcomp}
\usepackage{stfloats}
\usepackage{url}
\usepackage{verbatim}
\usepackage{graphicx}
\usepackage{cite}
\usepackage{wasysym}
\usepackage{makecell}
\usepackage{booktabs}

\begin{document}

\title{STEP: Detecting Audio Backdoor Attacks via Stability-based Trigger Exposure Profiling\thanks{This work has been submitted to the IEEE for possible publication. Copyright may be transferred without notice, after which this version may no longer be accessible.}}

\author{Kun Wang$^{1}$, Meng Chen$^{1}$, Junhao Wang$^{1}$, Yuli Wu$^{1}$, Li Lu$^{1,\ast}$, Chong Zhang$^{2}$,\\
Peng Cheng$^{1}$, Jiaheng Zhang$^{3}$, and Kui Ren$^{1}$\\
\small $^{1}$Zhejiang University \quad $^{2}$Xi'an Jiaotong University \quad $^{3}$National University of Singapore\\
\small $^{\ast}$Corresponding author: li.lu@zju.edu.cn
}


\maketitle

\begin{abstract}
With the widespread deployment of deep-learning-based speech models in security-critical applications, backdoor attacks have emerged as a serious threat: an adversary who poisons a small fraction of training data can implant a hidden trigger that controls the model's output while preserving normal behavior on clean inputs.
Existing inference-time defenses are not well suited to the audio domain, as they either rely on trigger over-robustness assumptions that fail on transformation-based and semantic triggers, or depend on properties specific to image or text modalities.
In this paper, we propose STEP (Stability-based Trigger Exposure Profiling), a black-box, retraining-free backdoor detector that operates under hard-label-only access.
Its core idea is to exploit a characteristic dual anomaly of backdoor triggers: anomalous label stability under semantic-breaking perturbations, and anomalous label fragility under semantic-preserving perturbations.
STEP profiles each test sample with two complementary perturbation branches that target these two properties respectively, scores the resulting stability features with one-class anomaly detectors trained on benign references, and fuses the two scores via unsupervised weighting.
Extensive experiments across seven backdoor attacks show that STEP achieves an average AUROC of 97.92\% and EER of 4.54\%, substantially outperforming state-of-the-art baselines, and generalizes across model architectures, speech tasks, an open-set verification scenario, and over-the-air physical-world settings.
\end{abstract}


\section{Introduction}

\begin{figure}[t]
  \centering
  \includegraphics[width=\columnwidth]{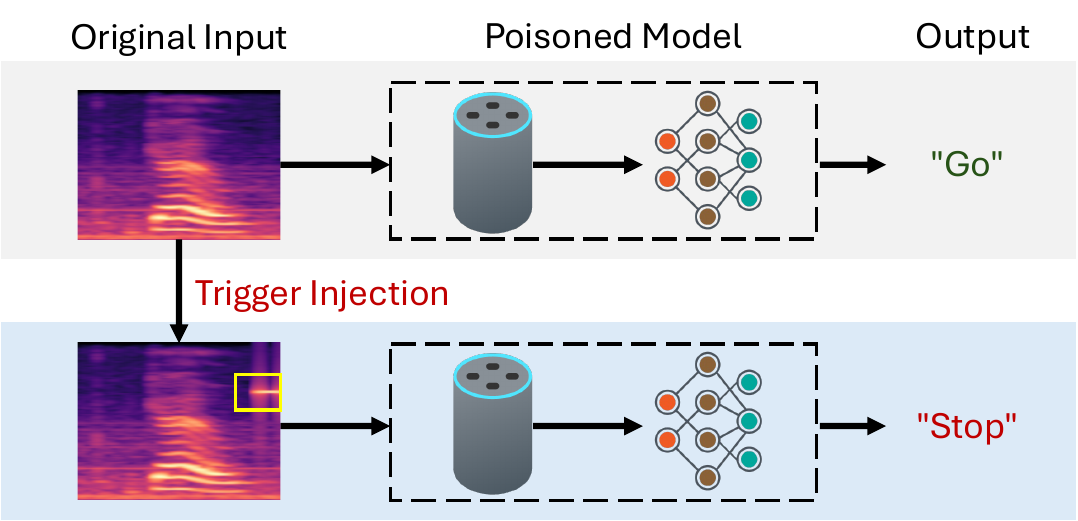}
  \caption{Illustration of a backdoor attack on a speech model. During training, a small fraction of samples are poisoned with a trigger pattern and relabeled to a target class. At inference time, the backdoored model behaves normally on clean inputs but produces the attacker-chosen output whenever the trigger is present.}
  \label{fig:attack_illustration}
\end{figure}

Driven by the rapid advancement of deep learning, speech models have achieved remarkable progress in tasks such as speaker recognition, speech command recognition, and automatic speech recognition, and are now widely deployed in security-critical applications, such as voice-activated assistants, speaker verification for financial authentication, and in-vehicle voice control systems.
The global speech and voice recognition market, valued at USD 19.09 billion in 2025, is projected to reach USD 104.05 billion by 2034 at a CAGR of 20.30\%~\cite{speech_market}, reflecting the increasing reliance on these systems in everyday life.
However, this widespread adoption also introduces new security risks.
In outsourced training and machine-learning-as-a-service (MLaaS) scenarios, an adversary can corrupt a model by poisoning only a small fraction of its training data, implanting a hidden backdoor trigger that causes the model to produce an attacker-chosen output on demand while behaving correctly on clean inputs, as illustrated in Fig.~\ref{fig:attack_illustration}.
Recent studies have demonstrated backdoor attacks on speech systems from multiple dimensions, including diverse trigger designs~\cite{koffas_dynamic_2022,Zhai_Backdoor_icassp2021,xin_natural_2023,chen_devil_2024,schoof_emoback_2024,yao_emoattack_2024}, physical-world delivery~\cite{zhang_inaudible_2024,zheng_silent_2023}, and different attack surfaces~\cite{lan_flowmur_2024,li_cuckooattack_2025,zong_trojanmodel_2023}, underscoring the practical urgency of developing effective defense mechanisms.

Existing approaches to detecting backdoor attacks~\cite{wang_neural_2019,tran_spectral_2018,ma_beatrix_2022,xie_badexpert_2023,gao_strip_2019,guo_scale_2023,udeshi_model_2022,liu_detecting_2023} have achieved promising results, but often require access to model internals, the training dataset, or model retraining, making them impractical in real-world MLaaS deployments where the defender has no visibility into the model pipeline.
The few methods that operate under weaker assumptions generally exploit the prediction stability of backdoor triggers under input perturbations, but each relies on modality-specific or trigger-specific assumptions.
For example, STRIP~\cite{gao_strip_2019} assumes input-agnostic triggers, and NEO~\cite{udeshi_model_2022} assumes localized, patch-based trigger patterns, both of which do not generalize to the diverse trigger designs observed in the audio domain.
As a result, these methods suffer from significant performance degradation or even complete failure when applied to speech models.
To our knowledge, the audio domain still lacks an effective, universal, and practical backdoor detection method that operates under a fully black-box, hard-label-only setting, which is the standard interface exposed by most commercial prediction APIs.

Toward this end, our work aims to propose a practical backdoor detection method for speech models, which is effective under black-box and hard-label-only scenarios, universal across different trigger types, robust in physical-world settings, and portable across various model architectures and speech tasks.
To achieve these goals, we face the following challenges:
\textit{(1) Trigger diversity:} Backdoor triggers in the audio domain span a wide range of injection methods, including additive patterns, global signal transformations, and semantic modifications. In a defense scenario, the attack strategy is completely unknown to the defender, so the detection method must not rely on any assumption about the trigger type.
\textit{(2) Strict black-box:} In a fully black-box setting, the defender has no access to model internals or output logits. Only the predicted hard label is available per query, providing severely limited information from which to construct a detection signal.
\textit{(3) Multiple tasks:} The target model's task is unknown to the defender. The output may represent a speaker identity, a speech command, or a binary accept/reject decision in a verification scenario. The detection method must accommodate this diversity without task-specific design.

In this paper, we present STEP (Stability-based Trigger Exposure Profiling), a unified, black-box, retraining-free backdoor detector for speech models that addresses all three challenges.
To handle trigger diversity, we revisit the stability properties of backdoor triggers under perturbation. Beyond the trigger over-robustness under semantic-breaking perturbations exploited by prior work, we additionally identify trigger fragility under semantic-preserving perturbations, and unify both perspectives through two complementary perturbation branches to cover a broader range of attack types.
To operate under the strict black-box constraint, we adopt a one-class anomaly detection approach that learns the benign stability profile from reference samples under both perturbation families, and fuses the two branch scores via unsupervised inverse-variance weighting, requiring no malicious labels or output probabilities.
To support multiple speech tasks, we design the stability profile based solely on whether the predicted label changes under perturbation, rather than interpreting the label content itself. This label-flip formulation is agnostic to the output space, making the method directly applicable to classification, recognition, and verification tasks without modification.

Our contributions are summarized as follows:
\begin{itemize}
  \item We propose a unified backdoor detection method for speech models that captures both trigger over-robustness and trigger fragility through two complementary perturbation branches, covering a broader range of trigger types than any single-perspective approach.

  \item We implement STEP under black-box, hard-label-only assumptions with three key design choices: label-flip comparison to accommodate diverse output spaces across speech tasks, one-class anomaly detection trained solely on benign samples to eliminate the need for malicious labels, and unsupervised score fusion to combine the two branches without manual tuning.

  \item We conduct extensive experiments across seven backdoor attacks and four state-of-the-art baselines. STEP achieves an average AUROC of 97.92\% and EER of 4.54\%, substantially outperforming all baselines, and generalizes across model architectures, speech tasks, and over-the-air physical-world settings.
\end{itemize}

\section{Related Work}
\label{sec:related}

\begin{figure*}[t]
  \centering
  \subfloat[Clean]{\includegraphics[width=0.19\textwidth]{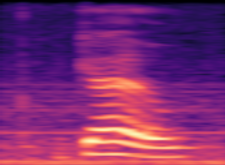}}\hfill
  \subfloat[SineTone]{\includegraphics[width=0.19\textwidth]{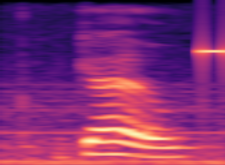}}\hfill
  \subfloat[Natural]{\includegraphics[width=0.19\textwidth]{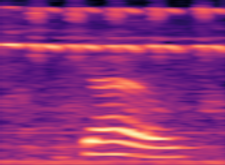}}\hfill
  \subfloat[JingleBack]{\includegraphics[width=0.19\textwidth]{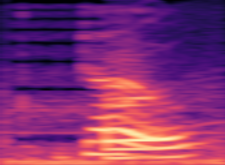}}\hfill
  \subfloat[TrojanRoom]{\includegraphics[width=0.19\textwidth]{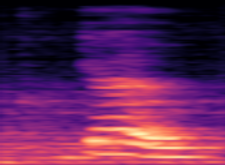}}
  \caption{Mel spectrograms of representative backdoor triggers. SineTone and Natural are additive triggers that inject localized signal patterns, whereas JingleBack and TrojanRoom apply global convolution-based transformations that alter the entire spectral structure yet remain imperceptible.}
  \label{fig:spectrograms}
\end{figure*}

\subsection{Backdoor Attacks on Speech Systems}
Backdoor attacks have been explored on a wide range of speech systems, including speech recognition \cite{lan_flowmur_2024, zong_trojanmodel_2023, li_cuckooattack_2025} and speaker recognition \cite{chen_devil_2024, yao_emoattack_2024, zhang_inaudible_2024}.
Based on the adversary’s capabilities, existing attacks can be broadly categorized into gray-box attacks, where the adversary can only poison training data, and white-box attacks, where the adversary additionally controls the training process or model architecture.

\subsubsection{Gray-box Attacks via Data Poisoning}
In gray-box settings, the adversary is assumed to be limited to poisoning a portion of the training dataset, without access to the model architecture or training procedure.
Early work explored simple signal-level triggers for backdoor attacks on speech systems.
Zhai et al. \cite{Zhai_Backdoor_icassp2021} proposed a low-amplitude one-hot spectrum noise as a trigger to attack speaker recognition models, while Koffas et al. \cite{koffas_dynamic_2022} introduced an equivalent sine-tone trigger in the audio domain to generate poisoned samples.
However, such triggers are often perceptible to human listeners, resulting in limited stealthiness.

To improve imperceptibility, subsequent studies investigated more subtle transformations as backdoor triggers.
Cai et al.~\cite{cai_toward_2024} proposed pitch boosting and sound masking (PBSM) as well as voiceprint selection and voice conversion (VSVC) to conceal noticeable patterns.
Similarly, Koffas et al. \cite{koffas_going_2023} introduced pitch-related stylistic transformations (JingleBack), while Yao et al. \cite{yao_imperceptible_2024} proposed a rhythm-based spectro-temporal transformation (RSRT) to further enhance imperceptibility.
Beyond prosody and rhythm manipulation, Schoof et al. \cite{schoof_emoback_2024} and Yao et al. \cite{yao_emoattack_2024} leveraged emotional voice conversion as triggers to attack speech and speaker recognition models.
More recently, PaddingBack \cite{ye_breaking_2024} and PhaseBack \cite{stealthy} explored padding-based and phase-injection hidden triggers to further reduce the perceptual footprint of poisoned samples.

While most prior work focuses on attacks in the digital domain, recent studies have begun to consider more realistic physical-world settings.
Inaudible ultrasonic signals \cite{zhang_inaudible_2024, zheng_silent_2023} and room impulse responses (TrojanRoom) \cite{chen_devil_2024} have been exploited as triggers to enable physical backdoor attacks, and MasterKey \cite{guo_masterkey_2023} demonstrated a universal backdoor trigger against speaker recognition systems. Figure~\ref{fig:spectrograms} illustrates the spectral characteristics of representative trigger types.
Building on these directions, FlowMur \cite{lan_flowmur_2024} proposed an optimized trigger learned on a surrogate model to improve attack transferability under limited knowledge, and CuckooAttack \cite{li_cuckooattack_2025} introduced an adaptive trigger injection mechanism based on a phoneme-level auxiliary dataset, enabling practical backdoor attacks against real-world automatic speech recognition systems.

\subsubsection{White-box Attacks via Model Control}
In white-box settings, the adversary is assumed to have full control over the training process and model architecture, enabling joint optimization of triggers and model parameters, or even direct modification of the model to implant architectural backdoors.
Under this stronger threat model, prior work has demonstrated powerful backdoor attacks on speech systems.

Early studies explored model-dependent trigger optimization and architectural backdoors.
TrojanNN \cite{liu_trojaning_2018} optimized carefully crafted background noise by manipulating intermediate representations, while TrojanNet \cite{tang_embarrassingly_2020} implanted explicit malicious modules into target networks to activate backdoors with simple triggers.
Subsequent work investigated more flexible trigger designs in the audio domain, including position-independent triggers \cite{shi_audio_2022}, passive backdoors exploiting ambient audio \cite{liu_opportunistic_2022}, and dynamic trigger generators jointly trained with the target model \cite{ye2022drinet}.
For automatic speech recognition, TrojanModel \cite{zong_trojanmodel_2023} demonstrated a practical white-box backdoor attack by jointly optimizing the acoustic model and trigger patterns, while stealthy frequency-domain trigger injection further improved imperceptibility \cite{zhang_stealthy_2024}.
Despite their effectiveness, such attacks rely on strong assumptions about adversarial access to training and deployment pipelines, limiting their applicability in many realistic scenarios.

\subsection{Backdoor Defenses}
Backdoor defenses aim to detect or mitigate backdoor behaviors, either before or during model deployment.
Based on the defender's access to training data and model internals, existing defenses can be broadly categorized into white-box, gray-box, and black-box approaches.

\subsubsection{White-box Defenses}
White-box defenses assume full access to both training data and model internals, enabling comprehensive intervention at any stage of the model lifecycle.
Training-stage defenses intervene during model optimization by purifying poisoned datasets \cite{tran_spectral_2018, hou_flare_2025} or constraining learning dynamics \cite{li_anti_2021, wang_training_2022, tang_setting_2023}, and have proven effective when the training pipeline is under the defender's control.
Post-training defenses operate after model training, focusing on offline auditing \cite{tang_demon_2021}, trigger inversion \cite{wang_neural_2019, wang_rethinking_2022, wang_unicorn_2023, xu_towards_2023}, statistical inspection \cite{qi_revisiting_2023, wang_mm_2024}, or model-level repair \cite{guan_backdoor_2024}, relying on white-box access to model parameters or intermediate representations. These methods have also been extended to non-standard learning paradigms such as competitive reinforcement learning \cite{guo_policycleanse_2023}. Beyond detection, some approaches further mitigate backdoor behaviors through neuron pruning \cite{liu_fine_2018}, adversarial unlearning \cite{zeng_adversarial_2022}, or activation correction \cite{li_purifying_2024, li_nearest_2024}, with extensions to model compression and large-scale representation learning \cite{zeng_beear_2024}. Despite their effectiveness, white-box defenses are fundamentally constrained by their strong access assumptions, limiting applicability in outsourced training or third-party deployment settings.

\subsubsection{Gray-box Defenses}
Gray-box defenses assume access to model internals such as intermediate representations or parameters, but not to the original training data. These methods typically operate at inference time, exploiting internal model signals to identify backdoor inputs without requiring retraining.
Several methods exploit representation-level or activation-space signals. BaDExpert \cite{xie_badexpert_2023} extracts backdoor-related functionality by probing the model and analyzing functional responses, enabling accurate detection without explicit trigger inversion. Beatrix \cite{ma_beatrix_2022} leverages high-order statistics of intermediate representations via Gram matrices, and is particularly effective against dynamic or sample-specific attacks. BadActs \cite{yi_badacts_2024} proposes a universal defense in the activation space by identifying abnormal activation patterns, while layer-wise feature analysis \cite{jebreel_defending_2023} highlights the broader utility of representation-based signals.
Beyond activation-level analysis, IBD-PSC \cite{hou_ibd_2024} leverages parameter-oriented scaling consistency to provide a robust inference-time detection signal. TED \cite{mo_robust_2024} takes a complementary perspective by tracking the layer-wise evolution of each input's neighborhood ranking, identifying backdoor inputs through anomalous evolutionary trajectories that persist even when single-layer feature separation fails. RBD \cite{yao_reverse_2024} detects backdoor inputs by comparing the deployed model against a distilled reference, exposing runtime behavioral discrepancies. REFINE \cite{chen_refine_2025} proposes an inversion-free defense via model reprogramming, suppressing backdoor behaviors at runtime without explicit trigger recovery. While effective, gray-box methods remain dependent on access to model internals, which may be unavailable in fully black-box deployment settings.

\begin{table}[t]
\centering
\small
\caption{Comparison of inference-time backdoor detection methods applicable to the audio domain. \CIRCLE/\Circle\ = satisfies/does not.}
\label{tab:defense_comparison}
\resizebox{\columnwidth}{!}{%
\begin{tabular}{lcccc}
\toprule
\textbf{Method}
& \textbf{Black-Box}
& \textbf{Retraining-Free}
& \textbf{Hard Label}
& \textbf{Universal} \\
\midrule
Beatrix \cite{ma_beatrix_2022}         & \Circle & \CIRCLE & \Circle & \CIRCLE \\
BaDExpert \cite{xie_badexpert_2023}    & \Circle & \Circle & \Circle & \CIRCLE \\
BadActs \cite{yi_badacts_2024}         & \Circle & \CIRCLE & \Circle & \CIRCLE \\
IBD-PSC \cite{hou_ibd_2024}           & \Circle & \CIRCLE & \CIRCLE & \CIRCLE \\
RBD \cite{yao_reverse_2024}           & \Circle & \Circle & \Circle & \CIRCLE \\
TED \cite{mo_robust_2024}             & \Circle & \CIRCLE & \Circle & \CIRCLE \\
\midrule
STRIP \cite{gao_strip_2019}           & \CIRCLE & \CIRCLE & \Circle & \Circle \\
NEO \cite{udeshi_model_2022}          & \CIRCLE & \CIRCLE & \CIRCLE & \Circle \\
SCALE-UP \cite{guo_scale_2023}        & \CIRCLE & \CIRCLE & \CIRCLE & \CIRCLE \\
TeCo \cite{liu_detecting_2023}        & \CIRCLE & \CIRCLE & \CIRCLE & \CIRCLE \\
\midrule
\textbf{STEP (Ours)}                  & \CIRCLE & \CIRCLE & \CIRCLE & \CIRCLE \\
\bottomrule
\end{tabular}
}
\vspace{2pt}
\raggedright\scriptsize
\textit{Black-Box}: no access to model internals;
\textit{Retraining-Free}: no retraining or fine-tuning of the target model required;
\textit{Hard Label}: only the predicted label required (no logits);
\textit{Universal}: no assumption on trigger type or attack strategy.
\end{table}

\subsubsection{Black-box Defenses}
Black-box defenses operate solely on model outputs, requiring neither training data nor access to model internals. This minimal assumption makes them the most broadly applicable class of defenses, particularly suited to outsourced training and third-party deployment scenarios where defenders have no visibility into the model pipeline.
A primary line of work exploits prediction consistency under input perturbations. STRIP \cite{gao_strip_2019} observes that input-agnostic backdoor triggers dominate model predictions even under strong superimposition perturbations, yielding abnormally low entropy for triggered inputs; this principle has since been generalized to multiple modalities \cite{gao_design_2022}. SCALE-UP \cite{guo_scale_2023} analyzes prediction stability under input scaling and identifies backdoor inputs through distinctive scaled-consistency patterns. TeCo \cite{liu_detecting_2023} further detects backdoor inputs by observing anomalous prediction robustness under systematic corruptions, while NEO \cite{udeshi_model_2022} demonstrates that output-level behavioral analysis generalizes across model architectures without model-specific assumptions.
Beyond classification, black-box defenses have been extended to generation-oriented tasks. ONION \cite{qi_onion_2021} filters suspicious tokens at inference time to defend against textual backdoor attacks, and BEAT \cite{yi_probe_2025} introduces a probing-based defense for large language models under extreme black-box constraints. Complementary studies further examine inference-time mitigation for natural language generation \cite{sun_defending_2023} and generative language models \cite{li_cleangen_2025}, as well as structured prediction tasks such as object detection \cite{zhang_test_2025}, demonstrating the broad applicability of output-level defense signals across tasks and modalities.

Despite this progress, as summarized in Table~\ref{tab:defense_comparison}, backdoor defenses for audio systems are still in their infancy. To the best of our knowledge, no existing defense has been specifically designed for audio backdoor detection under a black-box, hard-label-only setting. While some general-purpose black-box defenses such as STRIP \cite{gao_strip_2019} and SCALE-UP \cite{guo_scale_2023} can be directly applied, their effectiveness may diminish due to the unique acoustic properties of audio triggers, which differ fundamentally from visual or textual perturbations. This gap motivates our work.

\section{Problem Formulation}
\label{sec:problem}

\subsection{Task Definition}

We consider the problem of backdoor detection for deployed speech models at inference time.
Let $f: \mathcal{X} \rightarrow \mathcal{Y}$ denote a speech model, where $\mathcal{X}$ is the space of audio inputs and $\mathcal{Y}$ is a discrete output space (e.g., command labels in speech command recognition, speaker identities in speaker recognition, or transcription tokens in speech recognition).
The model $f$ may be backdoored: it behaves normally on clean inputs but produces an attacker-specified target output $y_t \in \mathcal{Y}$ whenever a trigger pattern is present in the input.

Given a test audio sample $x \in \mathcal{X}$ and query access to $f$, the goal of backdoor detection is to determine whether $x$ carries a backdoor trigger.
Formally, we seek a detector $D: \mathcal{X} \times \mathcal{F} \rightarrow \{0, 1\}$ such that:
\begin{equation}
D(x, f) =
\begin{cases}
1 & \text{if } x \text{ is a backdoor sample,} \\
0 & \text{otherwise.}
\end{cases}
\end{equation}
We focus on \emph{inference-time, input-level} detection: each sample is evaluated independently at test time, without modifying the model or accessing the training pipeline.
In this work, we instantiate and evaluate this framework on audio classification tasks, which represent the dominant setting in existing audio backdoor literature.

\subsection{Threat Model}

We characterize the defender's position in terms of deployment context, resource constraints, and prior knowledge.

\begin{itemize}
  \item \textit{Model access:} The defender interacts with the model as an external auditor without privileged access to the model provider's infrastructure. The model's architecture, parameters, and internal states (such as intermediate activations or gradients) are fully opaque to the defender. Only input-output queries are available, reflecting the typical interface exposed by third-party or machine-learning-as-a-service (MLaaS) deployments.

  \item \textit{Deployment constraints:} The defender encounters the model post-deployment and cannot intervene in the training pipeline. No retraining, fine-tuning, or architectural modification of the model is permitted; the defense must be applied to the model as-is, without altering its parameters or inference behavior.

  \item \textit{Output interface:} The deployment interface exposes only the predicted class label $\hat{y} = f(x)$ for each query, as is standard in most real-world prediction APIs. The defender has no access to the underlying output distribution, confidence scores, or any intermediate computation.

  \item \textit{Prior knowledge:} The defender has no prior knowledge of whether the model has been compromised, nor of the adversary's choice of attack strategy, trigger mechanism, or target class. No assumption is made about the trigger's form, scope, or origin; the defense must remain effective regardless of whether the backdoor was introduced via data manipulation, model modification, or any other means.
\end{itemize}

\subsection{Defense Objective}

We aim to design a backdoor detector that satisfies the following objectives.

\begin{itemize}
  \item \textit{Effectiveness:} The detector should reliably identify backdoor samples submitted to a compromised model, while maintaining a low false alarm rate on benign samples and benign models. This requires the detection signal to be both sensitive to backdoor behavior and robust against naturally occurring variations in clean inputs.

  \item \textit{Universality:} The detector should generalize across diverse backdoor attacks without being tailored to any specific trigger type or attack strategy. Detection should be constructed solely from benign reference samples, without relying on any knowledge of the trigger pattern, poisoning mechanism, or target label.

  \item \textit{Portability:} The detector should generalize across different model architectures and speech tasks (such as speech command recognition, speaker recognition, and speech recognition) with the same perturbation design and hyperparameters. The only per-model adaptation is re-estimating a reference distribution from a small set of benign samples of the target model, which is unsupervised and requires no knowledge of any attack.
\end{itemize}

\section{Proposed Method}
\label{sec:method}

\begin{figure*}[t]
  \centering
  \includegraphics[width=\textwidth]{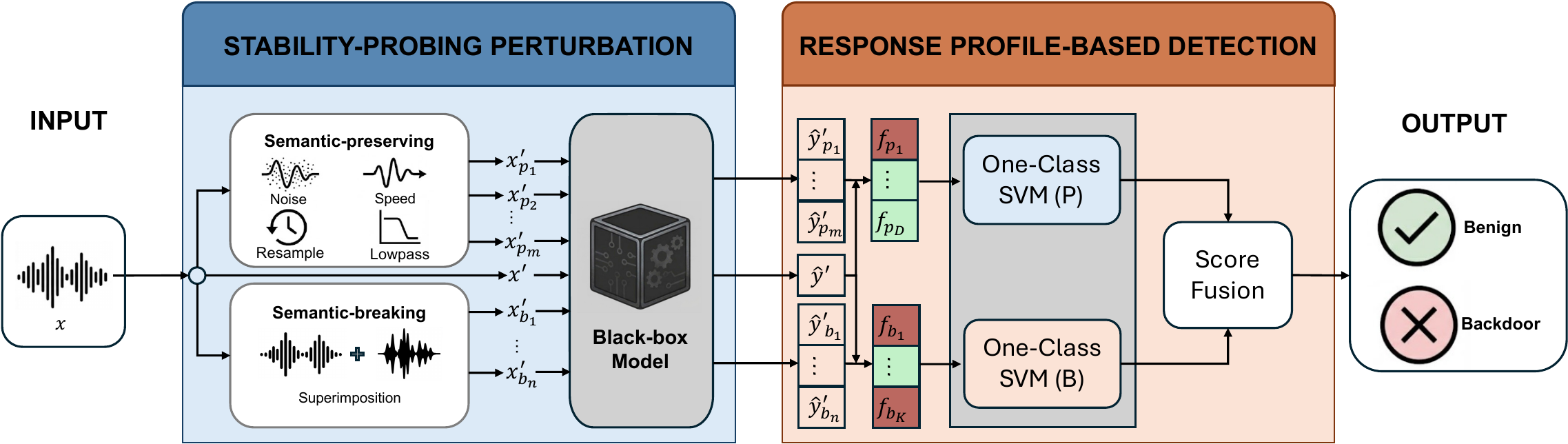}
  \caption{Overview of the STEP detection pipeline. The Stability-Probing Perturbation module applies semantic-preserving distortions and semantic-breaking superimpositions to each test input, producing a stability profile per family. The Response Profile-Based Detection module scores each profile with a one-class anomaly detector trained on benign references and fuses the two scores into a final detection decision.}
  \label{fig:overview}
\end{figure*}

\subsection{Overview}

Based on the design objectives established in Section~\ref{sec:problem}, we aim to build a backdoor detector that is effective, universal across attack types, and portable across model architectures and speech tasks, all under a black-box, hard-label-only setting that requires no retraining of the target model.

To this end, we propose STEP (Stability-based Trigger Exposure Profiling).
The core insight is that backdoor triggers, regardless of their specific design, differ fundamentally from genuine speech content in how they respond to perturbations.
STEP profiles prediction stability under two complementary perturbation types: \textit{semantic-breaking} perturbations that disrupt the task-relevant content of the utterance (e.g., mixing with another speech signal), and \textit{semantic-preserving} perturbations that alter acoustic conditions while keeping the semantic content intact (e.g., adding noise or changing the recording environment).
The combination of both perspectives enables STEP to cover a wider range of trigger types than any single-perspective defense.

Figure~\ref{fig:overview} illustrates the overall pipeline, which operates in two stages.
In the first stage, the Stability-Probing Perturbation module applies both perturbation families to each test input and records the model's prediction responses, producing a compact feature vector per family that summarizes the input's stability profile.
In the second stage, the Response Profile-Based Detection module trains a one-class anomaly detector on each feature space using only benign reference samples, scores each test input, and fuses the two scores into a single anomaly measure for the final detection decision.
The entire pipeline operates under hard-label queries only, requiring no access to model internals, output distributions, or the original training data.

\subsection{Design Motivation}

Existing black-box backdoor detection methods such as STRIP~\cite{gao_strip_2019} and SCALE-UP~\cite{guo_scale_2023} exploit the \textit{over-robustness} of triggers: because triggers dominate model predictions even under strong semantic-breaking perturbations, backdoored samples exhibit anomalously high prediction stability.
While effective for input-agnostic, patch-based triggers, this approach has a fundamental scope limitation.
Attacks employing transformation-based, environment-dependent, or frequency-localized trigger mechanisms may not survive semantic destruction, and thus evade over-stability detectors.

However, over-robustness is not the only exploitable property.
A complementary observation, supported by findings from TrojanModel~\cite{zong_trojanmodel_2023} and TrojanRoom~\cite{chen_devil_2024}, is that many triggers also exhibit \textit{fragility}: the gradient-optimized trigger in TrojanModel can be attenuated by basic signal processing, and the room-impulse-response trigger in TrojanRoom ceases to function when the acoustic environment changes.
Exploiting either property alone leaves blind spots for triggers that do not exhibit the targeted behavior.
STEP unifies both perspectives by combining semantic-breaking perturbations that expose trigger over-robustness with semantic-preserving perturbations that expose trigger fragility, remaining effective against additive, optimized, and environment-conditioned triggers within a single framework.

\subsection{Stability-Probing Perturbation}

The Stability-Probing Perturbation module applies two complementary perturbation families to the test input $x$, probing the model's prediction behavior from opposite directions.

\noindent\textit{Semantic-Breaking Perturbations:}
The first family mixes $x$ with a randomly drawn utterance $x_{\text{rand}}$ from the benign reference set at mixing coefficient $\alpha \in (0,1)$:
\begin{equation}
  \tilde{x} = \alpha \cdot x + (1-\alpha) \cdot x_{\text{rand}}.
\end{equation}
This operation is \textit{semantic-breaking} because it disrupts the task-relevant content of the utterance (linguistic content, speaker identity, and spoken command) proportional to $(1-\alpha)$.
For a benign sample, the model is forced to reconcile conflicting semantic signals from both utterances, yielding high prediction entropy and frequent label flips across repeated superimpositions.
A backdoored sample, by contrast, carries an implanted trigger that exerts disproportionate influence over the model's decision.
The trigger survives superimposition and continues to steer predictions toward the target class, producing anomalously low flip rates regardless of the mixed-in content.

We capture this signal through the \textit{flip curve}: for each of $K$ mixing coefficients $\{\alpha_1, \ldots, \alpha_K\}$, we compute the fraction of superimpositions (over multiple independently drawn $x_{\text{rand}}$) that change the model's top-1 prediction relative to the original.
This yields a $K$-dimensional feature vector $\mathbf{f}_{\text{B}}(x) \in [0,1]^K$ per sample, where continuous values capture the graded nature of semantic degradation across mixing levels.

\noindent\textit{Semantic-Preserving Perturbations:}
The second family applies acoustic transformations that alter recording conditions while leaving the task-relevant semantic content intact.
These perturbations are \textit{semantic-preserving} because a robust speech model should remain stable under natural acoustic variation: the spoken content, speaker identity, or spoken command is unaffected by recording environment or channel.
For example, we use additive noise, room impulse response convolution, low-pass filtering, resampling, and speed perturbation.

Backdoor triggers, by contrast, are typically engineered in a specific acoustic form and are fragile to such transformations: when the trigger is corrupted by distortion, the model's prediction reverts from the target class to the true class, producing a high label flip rate.
We apply $D$ fixed distortion configurations and record the binary flip indicator for each, yielding a $D$-dimensional binary feature vector $\mathbf{f}_{\text{P}}(x) \in \{0,1\}^D$ per sample. Under our hard-label-only setting, only the predicted label is observable, so each distortion produces a binary outcome: the prediction either flips or not.

\begin{algorithm}[t]
\caption{STEP Offline Training}
\label{alg:step_train}
\begin{algorithmic}[1]
\REQUIRE Model $f$, distortion set $\{d_1,\ldots,d_D\}$, mixing coefficients $\{\alpha_1,\ldots,\alpha_K\}$, benign reference set $\mathcal{D}_{\text{ref}}$, validation set $\mathcal{B}_{\text{val}}$
\ENSURE Detectors $g_{\text{P}}$, $g_{\text{B}}$, fusion weight $\beta^{*}$, threshold $\tau$
\FOR{each $x_i \in \mathcal{D}_{\text{ref}}$}
  \STATE $y_i \leftarrow f(x_i)$
  \FOR{$j = 1,\ldots,D$}
    \STATE $\mathbf{f}_{\text{P}}(x_i)_j \leftarrow \mathbf{1}[f(d_j(x_i)) \neq y_i]$
  \ENDFOR
  \FOR{$k = 1,\ldots,K$}
    \STATE Draw $N$ random utterances $\{x_{\text{rand}}^{(n)}\}_{n=1}^{N}$
    \STATE $\mathbf{f}_{\text{B}}(x_i)_k \leftarrow \frac{1}{N}\sum_{n=1}^{N} \mathbf{1}[f(\alpha_k x_i + (1\!-\!\alpha_k) x_{\text{rand}}^{(n)}) \neq y_i]$
  \ENDFOR
\ENDFOR
\STATE $g_{\text{P}} \leftarrow \textsc{Train}(\{\mathbf{f}_{\text{P}}(x_i)\}_{x_i \in \mathcal{D}_{\text{ref}}})$
\STATE $g_{\text{B}} \leftarrow \textsc{Train}(\{\mathbf{f}_{\text{B}}(x_i)\}_{x_i \in \mathcal{D}_{\text{ref}}})$
\STATE $\sigma_{\text{P}}^2 \leftarrow \operatorname{Var}_{x \in \mathcal{B}_{\text{val}}}[g_{\text{P}}(\mathbf{f}_{\text{P}}(x))]$; \quad $\sigma_{\text{B}}^2 \leftarrow \operatorname{Var}_{x \in \mathcal{B}_{\text{val}}}[g_{\text{B}}(\mathbf{f}_{\text{B}}(x))]$
\STATE $\beta^{*} \leftarrow \frac{1/\sigma_{\text{B}}^2}{1/\sigma_{\text{P}}^2 + 1/\sigma_{\text{B}}^2}$
\STATE Calibrate $\tau$ to target FPR on $\mathcal{B}_{\text{val}}$
\RETURN $g_{\text{P}}$, $g_{\text{B}}$, $\beta^{*}$, $\tau$
\end{algorithmic}
\end{algorithm}

\begin{algorithm}[t]
\caption{STEP Online Detection}
\label{alg:step_detect}
\begin{algorithmic}[1]
\REQUIRE Model $f$, test input $x$, detectors $g_{\text{P}}$, $g_{\text{B}}$, weight $\beta^{*}$, threshold $\tau$
\ENSURE $\delta(x) \in \{0,1\}$
\STATE $y \leftarrow f(x)$
\FOR{$j = 1,\ldots,D$}
  \STATE $\mathbf{f}_{\text{P}}(x)_j \leftarrow \mathbf{1}[f(d_j(x)) \neq y]$
\ENDFOR
\FOR{$k = 1,\ldots,K$}
  \STATE Draw $N$ random utterances $\{x_{\text{rand}}^{(n)}\}_{n=1}^{N}$
  \STATE $\mathbf{f}_{\text{B}}(x)_k \leftarrow \frac{1}{N}\sum_{n=1}^{N} \mathbf{1}[f(\alpha_k x + (1\!-\!\alpha_k) x_{\text{rand}}^{(n)}) \neq y]$
\ENDFOR
\STATE $s_{\text{P}} \leftarrow g_{\text{P}}(\mathbf{f}_{\text{P}}(x))$; \quad $s_{\text{B}} \leftarrow g_{\text{B}}(\mathbf{f}_{\text{B}}(x))$
\STATE Normalize $s_{\text{P}}, s_{\text{B}}$ to $\hat{s}_{\text{P}}, \hat{s}_{\text{B}} \in [0,1]$
\STATE $s_{\text{fused}} \leftarrow (1-\beta^{*})\,\hat{s}_{\text{P}} + \beta^{*}\,\hat{s}_{\text{B}}$
\STATE $\delta(x) \leftarrow \mathbf{1}[s_{\text{fused}} > \tau]$
\RETURN $\delta(x)$
\end{algorithmic}
\end{algorithm}

\subsection{Response Profile-Based Detection}

The Response Profile-Based Detection module learns the benign stability profile from reference samples, scores each test input for anomaly, and fuses the two branch scores into a final detection decision.

\noindent\textit{Anomaly Scoring:}
Since the defender has access only to benign samples, we cast detection as a one-class anomaly detection problem: a detector is trained on benign stability profiles and flags test samples that deviate significantly.
Let $\mathcal{D}_{\text{ref}}$ denote a held-out set of benign samples, split into a training portion and a validation portion $\mathcal{B}_{\text{val}}$ used for score normalization and threshold calibration.
For each perturbation family, we train a one-class detector on the benign reference feature vectors:
\begin{align}
  g_{\text{P}} &\leftarrow \text{train}\!\left(\{\mathbf{f}_{\text{P}}(x_i) \mid x_i \in \mathcal{D}_{\text{ref}}\}\right), \\
  g_{\text{B}}  &\leftarrow \text{train}\!\left(\{\mathbf{f}_{\text{B}}(x_i)  \mid x_i \in \mathcal{D}_{\text{ref}}\}\right).
\end{align}
Each detector produces an anomaly score per test sample: $s_{\text{P}}(x)$ and $s_{\text{B}}(x)$.
In our implementation, we instantiate $g$ as a one-class SVM with a linear kernel, though any one-class anomaly detector can be used.

\noindent\textit{Score-Level Fusion:}
Each score stream is normalized to $[0,1]$ using the benign validation percentiles, yielding $\hat{s}_{\text{P}}(x)$ and $\hat{s}_{\text{B}}(x)$, and combined as:
\begin{equation}
  s_{\text{fused}}(x) = (1-\beta)\,\hat{s}_{\text{P}}(x) + \beta\,\hat{s}_{\text{B}}(x),
\end{equation}
where $\beta \in [0,1]$ is set via unsupervised inverse-variance weighting on the benign validation set $\mathcal{B}_{\text{val}}$: a score stream with lower variance on benign samples indicates a tighter decision boundary and is assigned higher weight.
Concretely, we compute $\sigma_{\text{P}}^2 = \operatorname{Var}_{x \in \mathcal{B}_{\text{val}}}[\hat{s}_{\text{P}}(x)]$ and $\sigma_{\text{B}}^2 = \operatorname{Var}_{x \in \mathcal{B}_{\text{val}}}[\hat{s}_{\text{B}}(x)]$, and set:
\begin{equation}
  \beta^{*} = \frac{1/\sigma_{\text{B}}^2}{1/\sigma_{\text{P}}^2 + 1/\sigma_{\text{B}}^2}.
\end{equation}
This requires no malicious labels and is computable offline from the same benign reference data.

At test time, a sample is flagged as a backdoor input when $s_{\text{fused}}(x)$ exceeds a threshold $\tau$ calibrated to a target false positive rate on $\mathcal{B}_{\text{val}}$:
\begin{equation}
  \delta(x) =
  \begin{cases}
    1 \text{ (backdoor)} & \text{if } s_{\text{fused}}(x) > \tau, \\
    0 \text{ (benign)}   & \text{otherwise.}
  \end{cases}
\end{equation}
The detectors are trained once per target model using only benign samples, requiring no knowledge of the model's architecture, task type, or any attack. When deploying on a new target model, only the reference distribution needs to be re-estimated.
Algorithms~\ref{alg:step_train} and~\ref{alg:step_detect} summarize the complete offline training and online detection procedures, respectively.

\section{Experiments}
\label{sec:experiments}

\subsection{Experimental Setup}

\noindent\textit{Datasets:}
We evaluate on two benchmark datasets covering different speech tasks.
\begin{itemize}
  \item LibriSpeech (Libri)~\cite{panayotov_librispeech_2015}: a large-scale corpus of read English speech derived from LibriVox audiobooks, recorded at 16\,kHz. We use the \texttt{train-clean-100} subset ($\approx$100 hours) for training and the standard \texttt{test-clean} split for evaluation, formulating a closed-set speaker recognition (SR) task with 251 speakers.
  \item Google Speech Commands (GSC) v2~\cite{warden_speech_2018}: a speech command recognition (SCR) benchmark of $\approx$105,000 one-second utterances of 35 spoken words at 16\,kHz. We adopt the standard 12-class setting: ten target commands, plus silence and unknown-word classes.
\end{itemize}

\begin{table}[t]
\centering
\small
\caption{Attack reproduction results. ASR: attack success rate (\%); BA Drop: accuracy degradation (\%) from the clean model.}
\label{tab:attack_reproduction}
\begin{tabular}{llcc}
\toprule
\textbf{Attack} & \textbf{Trigger Type} & \textbf{ASR (\%)} & \textbf{BA Drop (\%)} \\
\midrule
SineTone   & Additive       & 99.20  & 0.52 \\
PBSM       & Spectral       & 95.80  & 1.05 \\
EmoBack    & Semantic       & 100.00 & 0.76 \\
JingleBack & Transformation & 97.00  & 0.58 \\
Natural    & Additive       & 100.00 & 0.52 \\
TrojanRoom & Transformation & 97.90  & 0.27 \\
Ultrasonic & Additive       & 100.00 & 0.99 \\
\midrule
Mean       &                & 98.56  & 0.67 \\
\bottomrule
\end{tabular}
\end{table}

\noindent\textit{Attacks:}
We reproduce seven backdoor attacks covering a broad range of trigger designs (Table~\ref{tab:attack_reproduction}):
SineTone~\cite{koffas_dynamic_2022} (fixed sine tone),
Ultrasonic~\cite{koffas_can_2022} (inaudible high-frequency tone),
Natural~\cite{xin_natural_2023} (environmental sound clip),
JingleBack~\cite{koffas_going_2023} (music jingle convolution),
TrojanRoom~\cite{chen_devil_2024} (room impulse response),
PBSM~\cite{cai_toward_2024} (spectral perturbation),
and EmoBack~\cite{schoof_emoback_2024} (prosodic modification).
All seven attacks achieve ASR above 95\% (mean 98.56\%) with BA drop below 1.1\% (mean 0.67\%), confirming effectiveness while preserving model utility.

\noindent\textit{Baselines:}
We compare STEP against four black-box inference-time defenses requiring no access to model internals or training data.
\begin{itemize}
  \item STRIP~\cite{gao_strip_2019}: superimposes test inputs with random clean samples and flags triggered inputs by their abnormally low prediction entropy. Unlike the other baselines, STRIP requires soft-label output probabilities rather than hard labels alone.
  \item SCALE-UP~\cite{guo_scale_2023}: detects backdoor inputs via their anomalously stable predictions under input amplitude scaling.
  \item NEO~\cite{udeshi_model_2022}: generates synthetic neighbors around each test input and identifies backdoor samples through anomalous neighborhood prediction distributions.
  \item TeCo~\cite{liu_detecting_2023}: exploits the observation that triggered inputs maintain suspiciously high prediction consistency under a suite of systematic corruptions.
\end{itemize}

\noindent\textit{Metrics:}
We evaluate detection performance using two complementary metrics.
\begin{itemize}
  \item AUROC (Area Under the Receiver Operating Characteristic Curve): measures the probability that a randomly chosen backdoor sample scores higher than a randomly chosen benign sample. As a threshold-free metric, AUROC summarizes discriminative ability across all operating points and is our primary comparison metric.
  \item EER (Equal Error Rate): the operating point at which false acceptance rate equals false rejection rate. A lower EER reflects better practical detection capability at the balanced threshold.
\end{itemize}

\noindent\textit{Implementation Details:}
The default target model is an x-vector network~\cite{snyder_x_2018} (xvect); portability experiments additionally use ECAPA-TDNN~\cite{desplanques_ecapa_2020}, CNN, and RNN architectures, and the realistic verification scenario uses a d-vector speaker encoder.
All experiments run on $2\times$ NVIDIA L40 GPUs (PyTorch 2.3.1, SpeechBrain 1.0.3).
All models take raw waveforms as input; to simulate realistic poisoning, triggers are injected prior to the training pipeline at a poison rate of 20\%, with standard data augmentation during training.
For the baselines: STRIP queries $N=10$ superimpositions and computes entropy over the output probability distribution; SCALE-UP scales amplitude across 5 levels from 1.2 to 2.0; NEO masks $N=10$ random 100\,ms segments with RMS-matched noise; TeCo applies 5 corruption types each at 5 severity levels.
For STEP: the semantic-preserving branch applies $D=11$ distortion configurations (additive noise, RIR convolution, dereverberation, resampling, low-pass filtering, and speed perturbation with varying parameter settings); the semantic-breaking branch uses $K=3$ mixing coefficients with 5 draws each (15 queries total).
All one-class SVM detectors are trained exclusively on benign samples using a linear kernel with $\nu=0.05$; the fusion weight $\beta$ is set via the unsupervised inverse-variance estimator throughout all experiments.

\subsection{Main Results}

\begin{table*}[t]
\centering
\small
\caption{Detection AUROC (\%$\uparrow$) of backdoor defenses on xvect + Libri (SR).
Bold indicates the best value per column.}
\label{tab:main_auroc}
\begin{tabular}{lcccccccc}
\toprule
\textbf{Defense} & \textbf{SineTone} & \textbf{PBSM} & \textbf{EmoBack} & \textbf{JingleBack} & \textbf{Natural} & \textbf{TrojanRoom} & \textbf{Ultrasonic} & \textbf{Avg.} \\
\midrule
STRIP    & \textbf{99.97} & 91.61 & 76.56 & 88.42 & \textbf{100.00} & 76.23 & 99.97 & 90.39 \\
NEO      & 48.91 & 49.82 & 50.16 & 50.17 & 50.14 & 50.44 & 47.27 & 49.56 \\
SCALE-UP & 50.05 & 50.00 & 50.01 & 50.00 & 50.00 & 50.01 & 49.99 & 50.01 \\
TeCo     &  8.82 & 16.40 & 13.02 & 21.04 & 10.34 &  4.77 & 85.36 & 22.82 \\
\midrule
STEP     & \textbf{99.97} & \textbf{97.34} & \textbf{97.65} & \textbf{94.96} & 99.85 & \textbf{95.67} & \textbf{100.00} & \textbf{97.92} \\
\bottomrule
\end{tabular}
\end{table*}

\begin{table*}[t]
\centering
\small
\caption{Detection EER (\%$\downarrow$) of backdoor defenses on xvect + Libri (SR).
Bold indicates the best value per column.}
\label{tab:main_eer}
\begin{tabular}{lcccccccc}
\toprule
\textbf{Defense} & \textbf{SineTone} & \textbf{PBSM} & \textbf{EmoBack} & \textbf{JingleBack} & \textbf{Natural} & \textbf{TrojanRoom} & \textbf{Ultrasonic} & \textbf{Avg.} \\
\midrule
STRIP    & 0.77  & 14.51 & 29.67 & 18.97 & \textbf{0.12}  & 30.19 &  0.52 & 13.54 \\
NEO      & 51.10 & 50.18 & 49.84 & 49.83 & 49.86          & 49.56 & 52.74 & 50.44 \\
SCALE-UP & 49.95 & 50.00 & 49.99 & 50.00 & 50.00          & 49.99 & 50.01 & 49.99 \\
TeCo     & 90.84 & 84.22 & 86.45 & 78.80 & 88.10          & 95.59 & 15.71 & 77.10 \\
\midrule
STEP     & \textbf{0.51}  & \textbf{8.47}  & \textbf{6.78}  & \textbf{8.13}  & 0.31  & \textbf{7.53}  & \textbf{0.06}  & \textbf{4.54} \\
\bottomrule
\end{tabular}
\end{table*}

Tables~\ref{tab:main_auroc} and~\ref{tab:main_eer} report detection AUROC and EER for all defenses on the xvect SR model trained on Libri, evaluated across seven backdoor attacks.

\textit{Baseline defenses:}
STRIP achieves strong detection on additive, input-agnostic triggers (SineTone: 99.97\%, Natural: 100.00\% AUROC) where triggers dominate predictions under superimposition, but degrades substantially on transformation-based and semantic attacks: AUROC drops to 76.56\% on EmoBack and 76.23\% on TrojanRoom, confirming that over-stability detectors fail when triggers do not survive semantic destruction.
SCALE-UP performs at chance level across all attacks (AUROC $\approx$ 50\%, EER $\approx$ 50\%).
This failure is rooted in a domain mismatch: the core assumption of SCALE-UP is that amplitude scaling amplifies trigger effects, which holds in image models that apply dataset-level global normalization.
Audio models, however, routinely apply per-sample normalization before inference, canceling out any amplitude change and rendering the scaling signal uninformative.
NEO likewise performs at chance level across all attacks.
NEO probes a sample by masking random 100\,ms temporal segments, implicitly assuming that triggers occupy a localized time region.
For the globally-applied triggers evaluated here (JingleBack, TrojanRoom, and EmoBack each transform the entire signal), no single masked segment removes the trigger effect, so backdoor and benign samples produce indistinguishable neighborhood prediction distributions.
TeCo exhibits inverted behavior (AUROC as low as 4.77\% on TrojanRoom and EER reaching 95.59\%), indicating that its corruption suite disrupts benign speech features more severely than it disrupts the evaluated triggers, causing the detection boundary to flip.

\textit{STEP:}
Our method achieves the best average AUROC of 97.92\% and average EER of 4.54\%, outperforming all baselines across both metrics.
Detection is near-perfect on additive triggers (SineTone: 99.97\% / 0.51\%, Ultrasonic: 100.00\% / 0.06\%) and remains strong on transformation-based attacks where baselines fail: TrojanRoom reaches 95.67\% AUROC / 7.53\% EER and EmoBack reaches 97.65\% / 6.78\%.
The only attack where STEP does not achieve the best EER is Natural, where STRIP obtains 0.12\% EER against 0.31\% for STEP; both are well within practical operating range and the gap in AUROC is negligible (100.00\% vs.\ 99.85\%).
These results demonstrate that unifying semantic-breaking and semantic-preserving perturbations under a single detection framework yields consistent coverage across diverse trigger types.

\subsection{Portability}

\begin{figure}[t]
  \centering
  \includegraphics[width=\columnwidth]{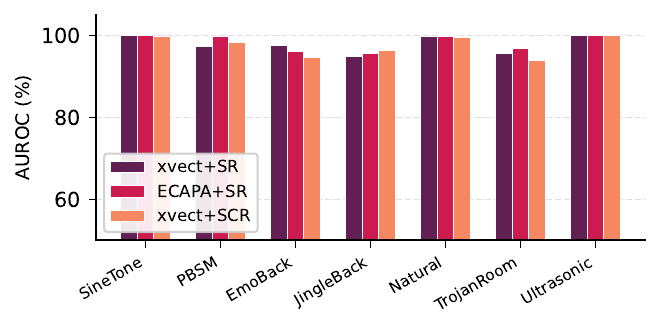}
  \caption{Portability across architectures and tasks: AUROC (\%$\uparrow$) under three settings. xvect+SR: main experiment; ECAPA+SR: architecture transfer; xvect+SCR: task transfer.}
  \label{fig:portability_auroc}
\end{figure}

\begin{figure}[t]
  \centering
  \includegraphics[width=\columnwidth]{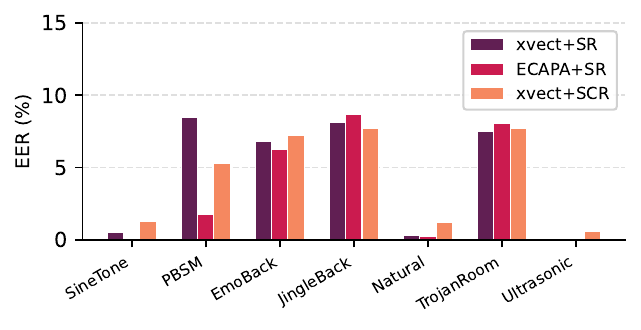}
  \caption{Portability across architectures and tasks: EER (\%$\downarrow$) under the same three settings.}
  \label{fig:portability_eer}
\end{figure}

Having established effectiveness across diverse attacks on a single model, we now examine whether the detection signal generalizes across model architectures, speech tasks, and their combination (Figs.~\ref{fig:portability_auroc} and~\ref{fig:portability_eer}; Table~\ref{tab:gen_arch_task}).
All perturbation configurations and detector hyperparameters remain identical to the main experiment; the only per-model adaptation is re-estimating the reference distribution from benign samples of the new target model.

\begin{table}[t]
\centering
\small
\caption{Average detection performance under simultaneous architecture and task transfer (SR $\to$ SCR).}
\label{tab:gen_arch_task}
\begin{tabular}{lcc}
\toprule
\textbf{Setting} & \textbf{\ AUROC (\%$\uparrow$)} & \textbf{\ EER (\%$\downarrow$)} \\
\midrule
xvect + Libri (SR) & 97.92 & 4.54 \\
CNN + GSC (SCR)            & 92.6  & 10.6 \\
RNN + GSC (SCR)            & 91.5  & 11.7 \\
\bottomrule
\end{tabular}
\end{table}

\begin{itemize}
  \item \textit{Across architectures:} Replacing xvect with ECAPA-TDNN on the same Libri SR task, STEP achieves 98.32\% AUROC and 3.59\% EER, marginally outperforming the xvect baseline (97.92\% / 4.54\%), confirming that the detection signal is architecture-agnostic.

  \item \textit{Across tasks:} Switching to an xvect SCR model on GSC changes both the task (SR $\to$ SCR) and the dataset (Libri $\to$ GSC). STEP achieves 97.5\% AUROC and 4.4\% EER, closely matching the main-experiment performance despite the joint shift.

  \item \textit{Across architectures and tasks:} Simultaneously changing both dimensions, we evaluate CNN and RNN SCR models on GSC. Average AUROC is 92.6\% (CNN) and 91.5\% (RNN), a moderate decline expected from the joint shift, while remaining well above chance level (Table~\ref{tab:gen_arch_task}).
\end{itemize}

These results confirm that the detection signal is not tied to a specific architecture or task: single-axis transfer preserves near-full performance, and even joint transfer retains strong detection, validating the portability objective in Section~\ref{sec:problem}.

\subsection{Realistic Verification Scenario}

The preceding experiments evaluate STEP on classification tasks. In practice, however, speech security systems predominantly operate as open-set verification systems.
To validate STEP in this more realistic setting, we deploy it on a speaker verification pipeline consisting of model training, speaker enrollment, and open-set verification via cosine similarity scoring with a threshold.
We use the Cluster attack~\cite{Zhai_Backdoor_icassp2021} with its publicly released codebase, targeting a d-vector speaker encoder trained on TIMIT.
The Cluster attack injects a one-hot spectrum pattern in the frequency domain, equivalent to a fixed sine-wave tone in the time domain, representing additive, input-agnostic triggers.

STEP achieves an AUROC of 99.8\% and EER of 1.2\%, confirming that the detection signal transfers to the open-set verification paradigm.
Notably, SCALE-UP, which fails at near-chance level in the main experiment, achieves above 98\% AUROC here due to the absence of per-sample normalization in the d-vector codebase. This further corroborates our earlier analysis that the failure of SCALE-UP is a normalization artifact rather than a fundamental limitation in the audio domain.

\subsection{Physical-World Evaluation}

\begin{figure}[t]
  \centering
  \includegraphics[width=0.8\columnwidth]{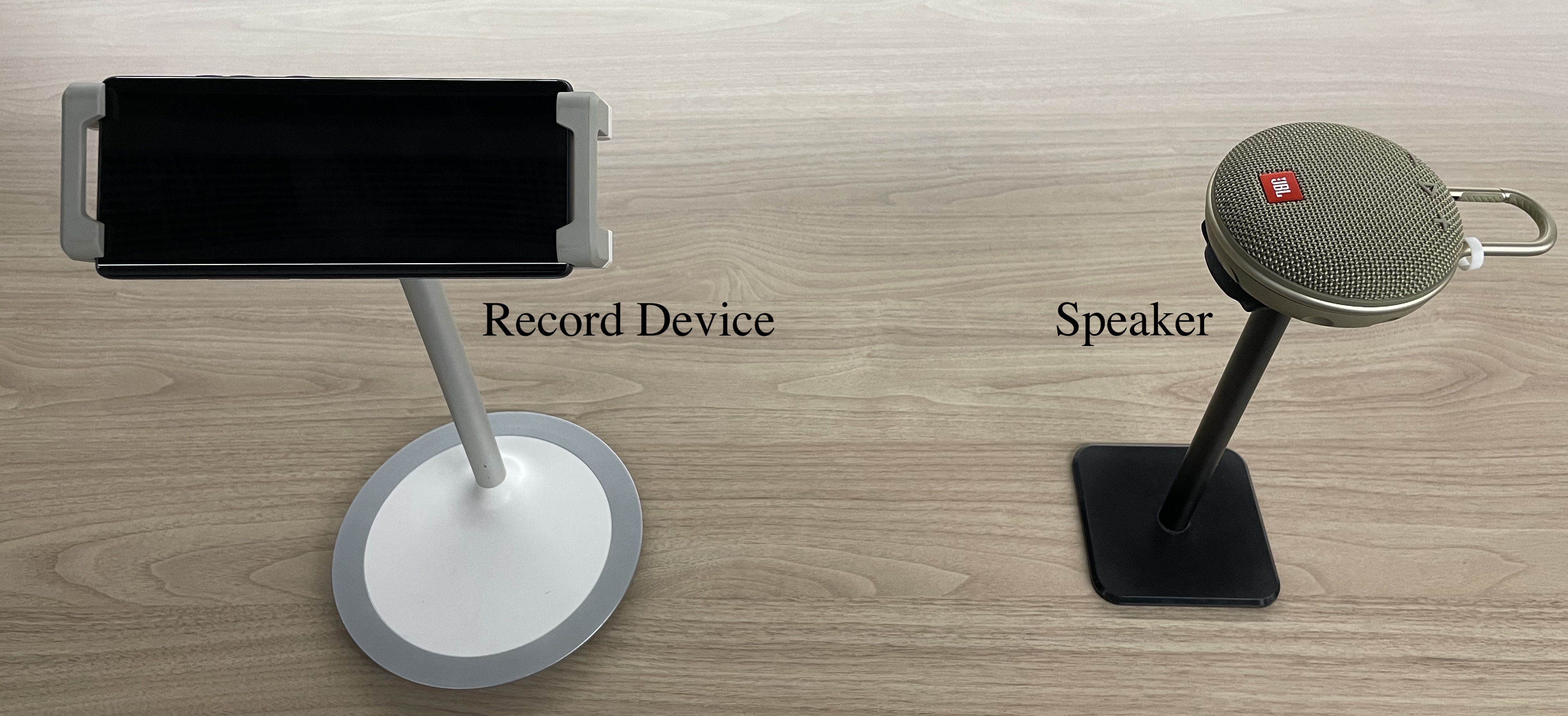}
  \caption{Physical-world recording setup. Backdoor samples are played via a loudspeaker and captured by three consumer devices in a meeting room.}
  \label{fig:physical_setup}
\end{figure}

\begin{figure}[t]
  \centering
  \includegraphics[width=\columnwidth]{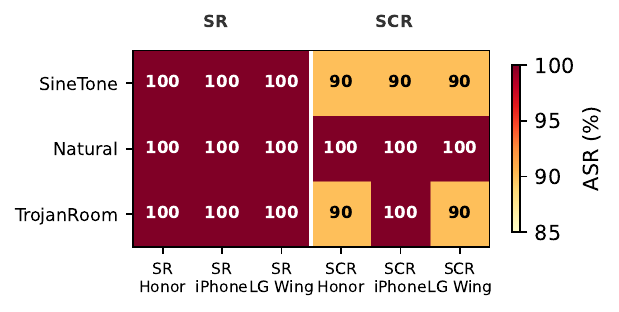}
  \caption{Physical-world ASR (\%) per attack, task, and device. STEP achieves 100\% detection on all attack-successful samples.}
  \label{fig:physical_results}
\end{figure}

We evaluate three physically realizable attacks (SineTone, Natural, and TrojanRoom) on xvect under both SR and SCR tasks.
For each attack and task, 10 backdoor samples are played back over a loudspeaker in a meeting room and re-recorded using three consumer devices: Honor Magic4 Pro, iPhone 12, and LG Wing, yielding 180 physically captured samples in total (3 attacks $\times$ 2 tasks $\times$ 3 devices $\times$ 10 samples).
Figures~\ref{fig:physical_setup} and~\ref{fig:physical_results} illustrate the recording setup and per-attack ASR results respectively.

On SR, all three attacks achieve 100\% ASR across all devices after physical delivery.
On SCR, Natural retains 100\% ASR, while SineTone drops to 90\% and TrojanRoom to 93.3\% on average, reflecting minor attenuation of high-frequency trigger components during acoustic transmission.
Despite this degradation, STEP achieves 100\% detection rate on all attack-successful samples across both tasks and all three devices, demonstrating that the perturbation-based detection signal is fully preserved under physical delivery.

\subsection{Ablation Study}

\begin{figure}[t]
  \centering
  \includegraphics[width=\columnwidth]{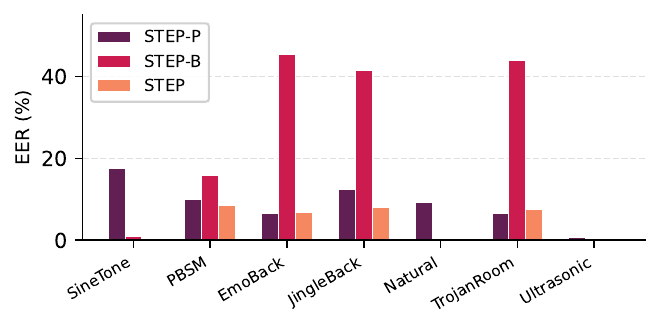}
  \caption{EER (\%$\downarrow$) of different STEP variants. STEP-P: semantic-preserving only; STEP-B: semantic-breaking only; STEP: with fusion.}
  \label{fig:ablation_eer}
\end{figure}

\begin{figure}[t]
  \centering
  \includegraphics[width=0.75\columnwidth]{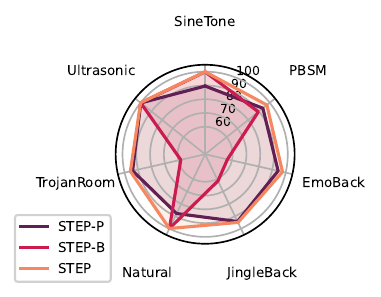}
  \caption{AUROC (\%$\uparrow$) of different STEP variants. STEP-P: semantic-preserving only; STEP-B: semantic-breaking only; STEP: with fusion.}
  \label{fig:ablation_radar}
\end{figure}

We ablate the contribution of each perturbation branch on xvect + Libri (Figs.~\ref{fig:ablation_eer} and~\ref{fig:ablation_radar}).

STEP-B alone achieves near-perfect detection on additive triggers (SineTone: 99.84\% AUROC / 0.94\% EER, Natural: 99.86\% / 0.30\%) but degrades sharply on transformation-based and semantic attacks (EmoBack: 56.61\% / 45.30\%, TrojanRoom: 58.13\% / 43.91\%).
STEP-P shows the complementary pattern: strong on transformation-based attacks (EmoBack: 94.37\% / 6.60\%, TrojanRoom: 94.04\% / 6.56\%) but weaker on additive triggers (SineTone: 89.52\% / 17.47\%).
As Fig.~\ref{fig:ablation_radar} illustrates, STEP-B exhibits a pronounced weakness on semantic and transformation-based attacks, while STEP-P is more uniform but falls short on additive triggers.
The fused system combines both branches and achieves consistent detection across all attack types, with an average AUROC of 97.92\% and EER of 4.54\%, outperforming either branch alone.

\section{Discussion}
\label{sec:discussion}

\textit{Computational overhead:}
STEP requires re-querying the target model for every perturbation configuration, totalling 26 forward passes per test sample (11 distortion configurations plus 15 superimposition draws).
In practice, all perturbations within each branch are independent and can be batched into a single forward pass, reducing wall-clock cost to that of a small batch inference rather than 26 sequential queries.
The anomaly scoring step is negligible in comparison.
Overall, STEP trades a modest increase in inference cost for substantially broader attack coverage, an acceptable overhead when security is the priority.

\textit{Adaptive attack resistance:}
An adversary aware of STEP could attempt to craft an adaptive trigger that evades both branches simultaneously: one robust enough to survive acoustic distortions (evading the semantic-preserving branch) while remaining inconspicuous under superimposition (evading the semantic-breaking branch).
Such joint evasion is non-trivial, as the two constraints impose conflicting requirements: a trigger robust to distortion must be pervasive in the signal, which tends to increase its dominance under superimposition and thereby make it more detectable by the semantic-breaking branch.
Formalising this trade-off and studying adaptive trigger designs that simultaneously satisfy both evasion objectives remains an important open problem for future work.

\textit{Portability to ASR systems:}
While the current evaluation focuses on classification and verification tasks, the detection principle extends naturally to automatic speech recognition (ASR) systems.
As demonstrated in TrojanModel~\cite{zong_trojanmodel_2023}, backdoor triggers in ASR can be effectively neutralised by standard acoustic preprocessing such as low-pass filtering and resampling, exhibiting the same trigger fragility that the semantic-preserving branch is designed to exploit.
The primary challenge lies in adapting the stability evaluation to sequence-level outputs: an ASR model produces a token sequence rather than a single label, so measuring prediction consistency requires sequence-level similarity metrics (e.g., character error rate or token-level agreement) in place of the binary flip indicator.
Extending STEP to this output paradigm is a natural direction for future work.

\section{Conclusion}
\label{sec:conclusion}

In this paper, we propose STEP, a black-box, retraining-free backdoor detection method for speech models that operates under hard-label-only access.
We observe that backdoor triggers exhibit characteristic dual anomalies: label stability under semantic-breaking perturbations and label fragility under semantic-preserving perturbations, and design a two-branch framework that profiles both behaviors and fuses the resulting anomaly scores via unsupervised weighting.
Extensive experiments demonstrate that STEP substantially outperforms existing baselines and generalizes across model architectures, speech tasks, an open-set verification scenario, and over-the-air physical-world settings.


\bibliographystyle{IEEEtran}
\bibliography{references}

\end{document}